# Collimation

*N. Fuster-Martínez*
Instituto de Física Corpuscular IFIC (CSIC-UV), Valencia, Spain

**Abstract**

Collimation systems are essential in particle accelerators to safely and efficiently manage unavoidable beam losses during operation. These systems rely on collimators, which are specially designed movable jaws or absorbers positioned close to the beam envelope to intercept and localize beam losses. Their role is particularly critical in high-intensity hadron machines, where uncontrolled losses can lead to equipment damage or operational downtime. While the specific requirements vary across accelerator types, circular accelerators, especially present and future high-energy colliders, cannot operate safely without a well-optimized collimation system. This lecture offers an overview of the fundamental principles, design challenges and operational strategies of beam collimation, with emphasis on high-intensity hadron accelerators. The Large Hadron Collider, the most advanced example to date, will serve as the main reference for illustrating state-of-the-art collimation approaches and technologies.

## 1 Introduction

Collimation is required to safely and efficiently manage unavoidable beam losses during operation. This is achieved using dedicated devices called collimators, typically fixed absorbers or movable jaws placed near the beam at optimized locations. The design and operational constraints of a collimation system depend on several key factors, including beam energy and intensity, the type of accelerator—linear or circular, normal-conducting or superconducting—and the facility's specific objectives. Accordingly, collimation systems are designed to fulfill multiple complementary roles, such as machine protection, beam loss control, preservation of beam quality, and reliable, efficient operation [1–3]. More specifically, the primary goals of a collimation system include:

– **Cleaning of betatron and off-momentum beam halo.** Beam halo particles with large transverse amplitudes or momentum deviations relative to the reference particle inevitably generate losses that must be intercepted and safely absorbed before reaching sensitive accelerator components. This is especially critical in superconducting machines, where even small losses can trigger magnet quenches. For example, at the Large Hadron Collider (LHC), the collimation system must ensure that local energy deposition remains below defined thresholds, on the order of a few tens of mW/cm³ at the superconducting dipoles [4, 5].

– **Passive machine protection.** Collimators form the first line of defense against beam losses under nominal, abnormal, or failure scenarios, protecting sensitive accelerator components. In modern high-intensity machines, where stored beam energies can be highly destructive, passive protection is essential. Consequently, machine protection is a key aspect of accelerator design [6, 7] and drives collimation system design, which must intercept losses, withstand credible failure scenarios, and ensure safe operation.

- **Radiation safety and activation control**. In high-power accelerators, the localization of beam losses in dedicated, well-shielded regions is becoming increasingly important to limit the long-term activation to a few areas in the accelerator. This facilitates safer and easier access for maintenance in the largest fraction of the accelerator that remains weakly or non-activated.

- **Control of collision debris and secondary particles in colliders**. In colliders, the reduction of uncontrolled irradiation of accelerator components caused by the products of beam–beam interactions is also essential, as such losses are unavoidable and may degrade the performance of other machine components and, ultimately, limit the achievable luminosity.

- **Background reduction in experiments**. Minimizing detector noise and spurious signals induced by beam halo particles is a key target performance goal for the collimation system of a collider, for both circular and linear machines. Avoiding beam halo particles scraping at the detector's location can reduce spurious signals in the detectors [8].

- **Beam shaping and emittance control**. Collimators are also used to improve the beam quality prior to high-precision sections, which is especially important in high-intensity machines (e.g. RFQs or downstream accelerator stages) or to tailoring the beam properties to meet specific goals.

- **Beam halo diagnostics**. Collimator scanning combined with sensitive beam loss monitors provides an effective way to probe the beam halo distribution. Collimators can also perform controlled beam scraping to shape tails and regulate intensity. This technique is routinely used in the SPS accelerator at CERN to optimize beam parameters before injection into the LHC and has been successfully applied in the LHC for experimental background control [9].

The collimation systems of modern high-power colliders have reached an unprecedented level of complexity. From the earliest high-energy circular colliders, such as the Tevatron [10], to the state-of-the-art LHC and its upcoming high-luminosity upgrade, the HL-LHC [11], the design and operation of collimation systems have become increasingly sophisticated. This evolution is driven by the pursuit of higher luminosity, $L$, which quantifies a collider's potential to enable new discoveries and depends critically on beam parameters such as the stored beam energy, $E_{stored}$, and the transverse beam size at the interaction point (IP) as:

$$L = \frac{1}{4\pi m_0 c^2} f_{rev} F \frac{N}{\beta^* \epsilon_n} E_{stored}, \qquad (1)$$

where $m_0$ is the particles rest mass, $c$ the speed of light, $f_{rev}$ the revolution frequency of the beam, $F$ is the geometric factor, $N$ is the bunch population, and assuming a round beam, $\beta^*$ is the betatron function at the IP in the transverse plane and $\epsilon_n$ is the transverse normalized emittance.

Equation (1) shows that two key parameters to enhance the luminosity performance of a collider are the increase of $E_{stored}$ and the decrease of the $\beta^*$. The collimation system plays a key role in the optimization of these two parameters. On the one hand, increasing the stored beam energy is only feasible if beam losses can be controlled safely and efficiently, ensuring that machine downtime remains acceptable and that accelerator components are not damaged. On the other hand, the collimation system must provide robust protection to the final-focus magnets, which are responsible for squeezing the beam to its minimum size at the IP. These magnets are among the most sensitive components of the accelerator, located in regions where the beam size reaches its maximum. As a result, in modern high-energy colliders, neither high-luminosity performance nor safe operation can be achieved without a well-designed and robust collimation system.

Figure 1 (left) shows a Livingston-like plot of the stored beam energy for past, present, and future high-power hadron and lepton accelerators. As illustrated, the LHC at CERN, originally designed to store up to 362 MJ and operated safely at about 430 MJ at 6.8 TeV, featuring stored beam energies exceeding those of its predecessors by nearly two orders of magnitude. Its high-luminosity upgrade, the HL-LHC, is expected to reach stored beam energies of approximately 700 MJ. Such extreme energy levels pose significant challenges, particularly in accelerators based on superconducting magnet technology. In these machines, even small uncontrolled beam losses can trigger magnet quenches, leading to beam dumps and increased machine downtime. In this context, the LHC collimation system represents the current state-of-the-art in beam loss control and machine protection. It comprises about 118 two-sided collimators strategically distributed around the 27 km ring, primarily in two dedicated collimation insertions. Figure 1 (right) shows the LHC ring with the two counter-rotating beams (beam 1 and beam 2), indicating the names and locations of the main LHC collimators in the two dedicated collimation sections IR3 and IR7. The interaction regions hosting the main particle physics experiments—ATLAS, ALICE, CMS, and LHCb—are also indicated.

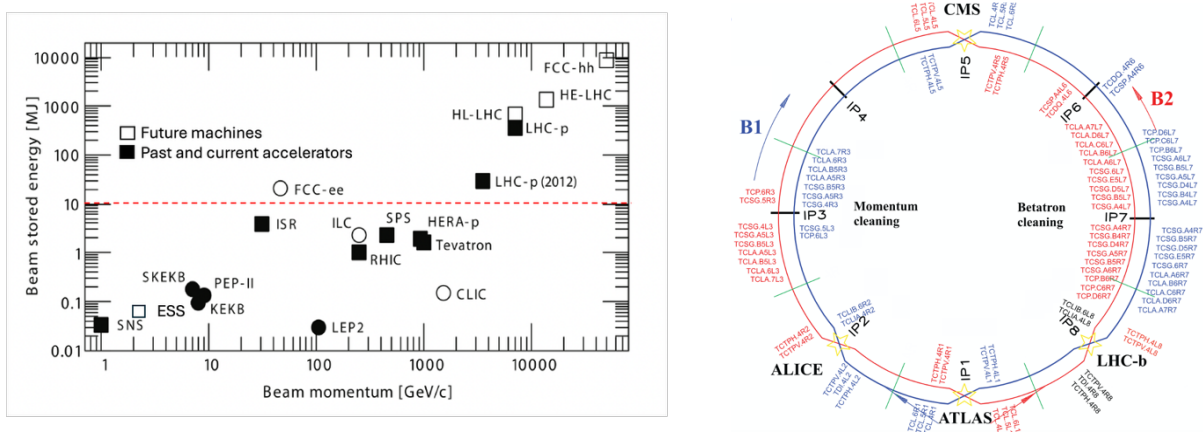

**Fig. 1:** Left: Livingston-like plot of stored beam energy for hadron and lepton high-power accelerators. Adapted from an initial version by R. Aßmann and S. Redaelli. Right: LHC collimation system for the two counterrotating beams [3].

## 2 Basic definition and inputs for collimation design

### 2.1 Beam halo

Hill's equation of motion describes the transverse dynamics of charged particle beams in an accelerator under the action of linear magnetic fields. In its homogeneous form, it governs the betatron motion of on-momentum particles and forms the basis for the definition of the fundamental optical functions of the lattice, such as the betatron functions $\beta(s)$ and the phase advance $\phi(s)$. In the context of collimation, however, it is necessary to account not only for particles with large betatron amplitudes (betatron halo), but also for particles with a significant momentum deviation with respect to the reference particle, $\Delta p/p_0$ (off-momentum halo). Including this effect leads to the non-homogeneous form of Hill's equation. Its general solution can be expressed as the sum of the homogeneous betatron solution and a particular solution proportional to the dispersion function *D(s)*, which also depends only on the lattice properties. *D(s)* quantifies the additional transverse displacement experienced by off-momentum particles along the accelerator. Incorporating this term, the general solution for the horizontal plane is:

$$x(s) = A\sqrt{\beta_x(s)}\cos(\phi_x(s) + \phi_0) + D_x(s) \times \frac{\Delta p_x}{p_0}, \qquad (2)$$

where *A* is a constant of motion related to the betatron oscillation amplitude, $\beta_x(s)$ is the horizontal betatron function describing the transverse focusing properties of the lattice, and $\phi_x(s)$ is the horizontal phase advance, specifying the evolution of the particle oscillation along the machine, $\phi_0$ a constant of integration, $D_x(s)$ is the horizontal dispersion function and $\Delta p_x/p_0$ the horizontal momentum deviation with respect to the reference particle. The same solution applies to the vertical plane by changing the *x* with *y* in Eq. (2).

At a given longitudinal position *s* along the accelerator, the transverse motion of a particle in phase-space *(x, px)* is represented by an ellipse. For a beam composed of an ensemble of particles with different initial conditions, each particle follows its own ellipse within this phase space. The transverse emittance of the beam is commonly defined as the area of the phase-space ellipse containing a specified fraction of the beam population, typically corresponding to 95%.

The projection of the phase space particles distribution onto the horizontal axis corresponds to the horizontal beam size as:

$$\sigma_x = \sqrt{\epsilon_x \beta_x(s) + (D_x(s)\delta_x)^2}, \qquad (3)$$

where $\delta_x$ corresponds to the rms bunch energy spread $(\Delta p_x/p_0)_{rms}$. A similar result can be obtained for the vertical plane by substituting *x* with *y* in Eq. (3). Equation (3) shows that the transverse beam size consists of two distinct contributions, leading to two types of beam halo that must be collimated: the betatron halo and the off-momentum halo. A typical transverse halo can be defined as the fraction of particles extending beyond about 3σ, where σ denotes the transverse beam size, representing the outermost population surrounding the beam core. This concept is illustrated in Fig. 2, which displays a normalized Gaussian beam-core distribution (red) with overpopulated tails (blue) intercepted by collimators (black rectangles). The transverse amplitude is expressed in units of beam size.

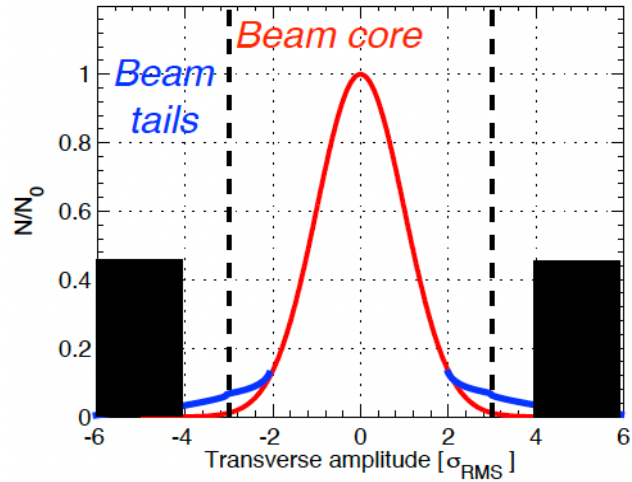

**Fig. 2:** Normalized Gaussian beam-core distribution (red) with overpopulated tails (blue) to be intercepted by collimators illustrated with black rectangles [2]. The transverse amplitude is given in units of beam size.

These concepts are essential for determining the optimal placement of collimators, as they dictate where beam amplitudes are largest and where beam halo particles can be most effectively intercepted to control losses and protect the machine. Therefore, to effectively address both components, collimators should ideally be placed at locations with zero dispersion for efficient betatron cleaning, while regions with large dispersion and low betatron amplitude are best suited for off-momentum cleaning.

## 2.2 Normalized aperture, aperture model and bottleneck

Because the beam size varies along the machine, the *normalized aperture*, $A_{x,y}$, is typically used to define machine apertures and collimator settings. $A_{x,y}$ is defined as the geometrical aperture of the accelerator components, $r_{x,y}(s)$, divided by $\sigma_{x,y}(s)$, as:

$$\boldsymbol{A_{x,y}} = \frac{r_{x,y}(s)}{\sigma_{x,y}(s)}. \tag{4}$$

Equation (4) provides a dimensionless location-independent measure of how close a physical aperture is to the beam core, independent of the local optical functions. This normalization allows apertures at different locations along the accelerator to be directly compared.

In high-energy circular colliders, the circulating beams must never directly intercept the beam pipe or other machine components except for some collimators; therefore, the location exhibiting the smallest normalized aperture in the machine, commonly referred to as the aperture *bottleneck*, must be reliably protected by the collimation system. Identifying and quantifying this bottleneck requires the construction of a detailed aperture model of the accelerator accounting for the complete geometry of all machine elements—including their shape, dimensions, position, and orientation—as well as the beam orbit, optical functions (betatron and dispersion), manufacturing tolerances, alignment uncertainties, and other machine imperfections.

Dedicated tools have been developed for this purpose and are integrated into optics and tracking codes such as MAD-X [12] and SixTrack [13], enabling a systematic evaluation of the machine aperture while consistently accounting for these effects. These tools have been extensively used both during the design phase and in operational support of the LHC, allowing the definition of robust and reliable collimation settings. Over the years, their predictions have been systematically benchmarked against experimental measurements, thereby increasing confidence in the underlying methodology [14].

## 2.3 Beam loss model

Beam losses are unavoidable and arise from a variety of physical processes and operational conditions. Depending on their origin, beam losses can be broadly classified into regular and abnormal losses. Regular losses are intrinsic to standard machine operation and include effects such as interactions with residual gas, intra-beam scattering, beam instabilities (single-bunch, collective, and beam-beam effects), dynamic changes during the operational cycle (e.g. orbit drifts, optics variations, and energy ramps), transverse resonances, RF noise, as well as losses associated with injection, beam dumping, and particle burn-off at the collision points in colliders. In contrast, abnormal losses originate from failures of accelerator subsystems or from incorrect beam manipulation, including human error.

Understanding and modelling these different loss mechanisms is essential for the design and optimization of effective collimation systems and forms the foundation of a comprehensive beam loss model. In many cases, accurately modelling beam losses requires detailed simulations that account for radiation–matter interactions, secondary particle production, and the modelling of the beam halo distribution in the machine under study. However, for circular accelerators, and when considering only losses arising from diffusion-like mechanisms, a simplified model can be constructed without explicitly simulating all these effects which can be used in early design phases. Under these conditions, the beam losses can be approximated by an exponential decay function characterized by a finite, time-dependent parameter known as the beam lifetime, $\tau_b(t)$, as:

$$I(t) = I_0 e^{-\frac{t}{\tau_b(t)}}, \qquad (5)$$

where $I_0$ is the initial beam intensity and $\tau_b(t)$ the beam lifetime representing the time after which the beam intensity is reduced to 1/e (approximately 37%) of its initial value. Note that the beam lifetime in Eq. (5) cannot be assumed to be constant throughout the operational cycle, since different loss mechanisms may dominate at different stages of machine operation. For a given beam lifetime, the corresponding loss rate can be calculated as:

$$-\frac{1}{I_0}\frac{dI}{dt} = \frac{1}{\tau_b(t)}. \qquad (6)$$

The value computed using Eq. (6) gives us an estimate of the beam losses that the collimation system must be able to intercept. Beam lifetime assumptions can be combined with particle tracking simulations and sensitivity studies to ensure that, even in scenarios with rapid losses (short $\tau_b(t)$), the designed collimation system can absorb the resulting radiation loads. This approach was used in the early LHC collimation design phase. Assuming a pessimistic beam lifetime value of 0.2 hours at nominal intensity, the loss rate is of approximately 4.5 x $10^{11}$ protons per second. At a proton energy of 7 TeV, this translates into a beam loss power of about 500 kW.

## 2.4 Cleaning efficiency

The performance of a collimation system can be measured in terms of the *cleaning inefficiency*, $\eta_C$, which reflects how well the collimation system prevents unwanted beam losses in sensitive regions. It is defined as the ratio of particles lost in sensitive regions, $A_{lost}$, to those absorbed by the collimators, $A_{coll}$ as:

$$\eta_C = \frac{A_{lost}}{A_{coll}}. \qquad (7)$$

Defining a collimation cleaning inefficiency criterion depends on the machine and application and can be different at different locations (e.g. quench limit of superconducting magnets, backgrounds at the experiments, activation…). Therefore, it is useful to define the local cleaning inefficiency, $\tilde{\eta}_C(s)$, defined as a function of the longitudinal coordinate $s$ as a fractional loss per unit length expressed in units of 1/m as:

$$\tilde{\eta}_C(s) = \frac{N(s \rightarrow s+\Delta s)}{N_{abs}} \frac{1}{\Delta s}, \qquad (8)$$

where $N(s \rightarrow s + \Delta s)$ is the number of particles lost over a length $\Delta s$ and $N_{abs}$ is the number of particles absorbed by the collimation system. Note that in Eq. (8) the longitudinal distribution must be evaluated over an adequate $\Delta s$ depending on layout considerations, e.g., over the length of a magnet. This definition allows direct comparison with the quench limit of the superconducting magnets. Currently, detailed and increasingly precise simulation tools are available to estimate the energy deposition in magnet coils, allowing direct comparison with the limits of superconducting cables. While such advanced tools are essential for detailed studies, this definition remains particularly useful during the initial layout and conceptual design phases.

## 3 Collimation design aspects for high-power machines

The design of a collimation system for high-power accelerators is driven by several critical factors, including the stored beam energy, expected loss scenarios, machine aperture constraints, required cleaning inefficiency by sensitive components, material robustness, and impedance-induced beam dynamics effects. These aspects must be addressed within a coherent design framework that ensures both safe operation and high machine performance.

The design workflow progressively refines both the conceptual layout and the technical implementation. In an initial phase, the goal is to establish a global understanding of the machine aperture hierarchy and cleaning strategy, based on key inputs such as aperture bottlenecks, sensitive components and tolerable loss rates, target values for local and global cleaning inefficiencies, and simplified beam loss models. With this information a preliminary baseline collimation layout is defined, specifying the number of collimators, type, longitudinal positions, and collimation depth of the collimators, together with preliminary material choices. When possible, the optics of the collimation section may also be designed or adjusted to optimize betatron and dispersion functions for efficient betatron or off-momentum cleaning. In a second phase, the proposed layout is validated through detailed tracking simulations to assess its performance, Monte Carlo simulations to evaluate energy deposition and secondary radiation, collimator material response and beam dynamics studies to quantify the impact of collimators on impedance and overall machine stability. The results of these simulations often feed back into the layout, optics, or material selection, forming an iterative optimization loop. Once satisfactory performance is confirmed, the design proceeds to detailed engineering, including mechanical integration, cooling, vacuum systems, and a final assessment of impedance contributions.

One of the key design aspects in high-power machines is the adoption of multi-stage collimation instead of a single-stage scheme, despite the significant increase in system complexity. While single-stage collimation may be sufficient for low-power accelerators, multi-stage collimation is essential in high-power machines, where extremely high cleaning efficiencies are required, as discussed in the following section.

### 3.1 Single vs multi-stage collimation system

A single-stage collimation system is composed of a single primary collimator (TCP). Such a collimator would ideally be installed in a warm region of the accelerator, as far as possible from superconducting magnets. Its jaws would be positioned at a normalized transverse aperture smaller than that of the machine bottleneck, thereby defining the limiting aperture of the machine. However, this concept would only be viable if the primary collimator acted as an ideal *black absorber*, capable of stopping all halo particles in a single passage through the jaw material. In practice, this assumption does not hold. For typical loss mechanisms in circular accelerators, halo particles reach the collimators with extremely small impact parameters, often in the sub-micrometre range. At such scales, not all particles traverse the full jaw length, particularly when unavoidable surface roughness and flatness errors are considered. Therefore, many particles experience only partial interactions—such as multiple Coulomb scattering or single diffractive processes—and continue circulating in the machine with increased transverse angles. These particles form what is commonly referred to as a *secondary halo*, which can be intercepted by other accelerator components before returning to the TCP. In addition, the interaction of the beam with the primary collimator material gives rise to hadronic and electromagnetic particle cascades that are not fully contained within the primary collimator itself. Secondary particles emerging from these showers can propagate downstream and reach sensitive components, such as superconducting magnets or beam instrumentation, further increasing the risk of quenches or radiation damage. These intrinsic limitations render single-stage collimation inadequate for high-power accelerators and naturally motivate the adoption of multi-stage collimation schemes.

Figure 3 illustrates the scheme of a single-collimation system (left) and the cleaning inefficiency (right) computed using SixTrack [13] for beam 1 around the ring for a configuration with only a single primary collimator in IR7. In this representation, peaks correspond to losses occurring at collimators (black), cold magnets (blue) and warm magnets (red). The results clearly show that, at several locations around the ring, near the physics interaction regions, IR5 and IR1, losses in cold elements reach cleaning inefficiency levels of the order of 0.001/m. These values exceed the maximum tolerable loss levels of superconducting magnets to avoid quenches. It should be noted that this estimate considers only primary particle losses in an idealized machine, excluding additional energy deposited by hadronic and electromagnetic showers, which further degrades the realistic scenario.

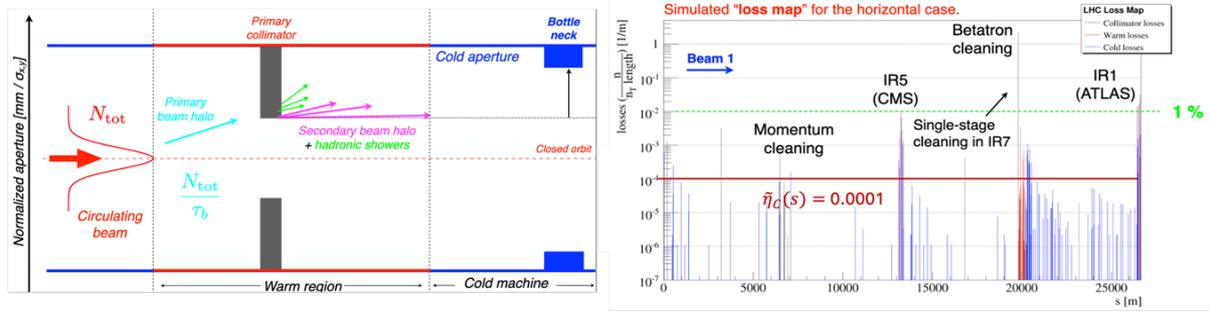

**Fig. 3:** Single-stage collimation scheme and loss map example. Left: Single-collimation scheme with the primary collimator and bottleneck illustrated. Right: Cleaning inefficiency in the LHC ring with loss peaks shown in blue at cold magnets, red at warm magnets and black at collimators. The local required cleaning inefficiency at the superconducting magnets is indicated with a red line. Thanks to D. Mirarchi and S. Redaelli.

A multi-stage collimation system enhances the performance of a single-stage scheme by introducing additional collimators downstream of the primary collimators to intercept secondary halo particles and the products of beam–matter interactions. In this approach, TCPs define the smallest normalized aperture in the machine and act as controlled scattering sources, while secondary collimators (TCSs), typically longer to maximize absorption, are placed at slightly larger apertures to capture the scattered particles. The system performance relies on a well-defined collimation hierarchy, usually satisfying that the losses at the TCP are larger than at the TCSs which are larger than those at the bottleneck ($n_{\text{TCP}} > n_{\text{TCS}} > n_{\text{BTNK}}$). In addition to aperture settings, the longitudinal placement of secondary collimators is optimized according to the phase advance with respect to the primary collimators, such that particles scattered at the TCPs reach maximum transverse offsets at the TCS locations. Detailed optimization of the phase advance settings for a two-stage collimation system in the presence of two-dimensional betatron halos can be found in Ref. [15]. It can be shown that an arrangement of primary and secondary collimators distributed over three planes—horizontal, vertical, and skew—allows for efficient multi-turn cleaning and provides satisfactory overall performance. The LHC collimation design can be found in Ref. [16].

While a two-stage collimation scheme—based on primary and secondary collimators—can efficiently shield the machine aperture from betatron and off-momentum halo losses, it is generally insufficient to fully absorb the products of hadronic and electromagnetic showers generated during particle–matter interactions. In addition, a two-stage system localized in a limited region of the accelerator does not provide adequate local protection for critical aperture bottlenecks that may be distributed along the machine, particularly in regions where optics changes, enhance beam sensitivity and in case of failure scenarios. These limitations naturally lead to the concept of a *multi-stage collimation system*, in which additional collimators are deployed with distinct but complementary roles. Beyond primary and secondary collimators, tertiary collimators (TCTs) are introduced to provide local protection of sensitive elements located close to the beam, while dedicated absorbers are used to intercept and contain secondary particle showers, reducing radiation loads and energy deposition in downstream components as illustrated in Fig. 4 left. This distributed, hierarchical approach allows both efficient halo cleaning and controlled management of secondary radiation, which are essential requirements for high-power accelerators. As an illustrative example, the collimation system of the LHC implements such a multi-stage concept, combines TCP, TCS and TCT collimators together with dedicated absorbers (TCLAs), and comprises a total of about 118 collimators distributed around the machine for both counter-rotating beams. The results of an LHC multi-stage collimation system loss map simulated with SixTrack is shown in Fig. 4 right. As shown in this loss map example, the blue loss peaks along the ring remain safely below the superconducting quench threshold.

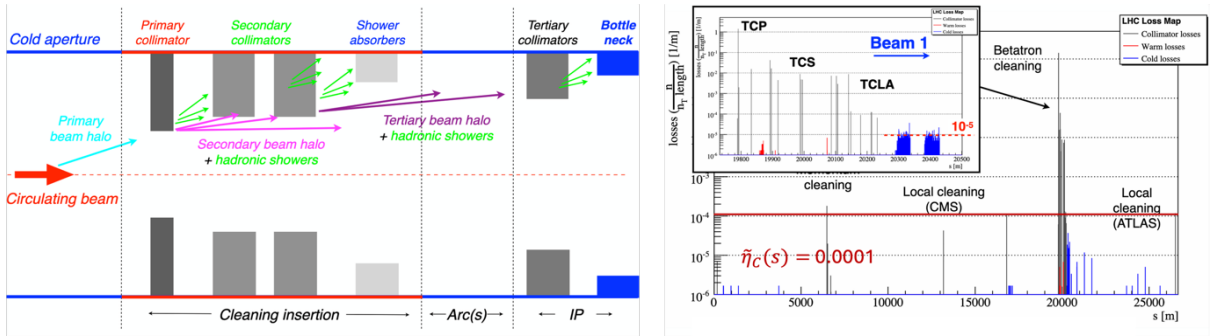

**Fig. 4:** LHC multi-stage collimation system scheme and LHC loss map. Left: LHC multi-stage collimation scheme. Right: Cleaning inefficiency in the LHC ring simulated for beam 1. Loss peaks shown in blue at cold magnets, red at warm magnets and black at collimators. The local cleaning inefficiency at the superconducting magnets is indicated with a red line. Thanks to D. Mirarchi and S. Redaelli.

The development of advanced simulation tools has been essential for the optimization and validation of modern multi-stage collimation systems such as the one implemented in the LHC. Achieving reliable performance predictions in these machines requires exceptional levels of physical and numerical accuracy. Modern high-energy accelerators incorporate many collimators made of different materials, and halo particles typically impact the collimator jaws with sub-micrometre impact parameters. At the same time, extremely high statistical precision is required to predict local cleaning inefficiencies at the level of 0.0001 or below. Such demanding conditions make collimation simulations particularly challenging. Accurate modelling further requires a detailed description of the machine aperture over kilometre-scale lattices, precise multi-turn tracking of large-amplitude halo particles, and the inclusion of relevant machine imperfections such as alignment errors, jaw tilts, surface roughness, and optics perturbations. Only by consistently accounting for all these effects it is possible to compute realistic performance estimates and establish safe operational margins.

To address these challenges, highly sophisticated and accurate toolchains have been developed over the past decades, combining multi-turn tracking codes such as SixTrack [13] with detailed particle–matter interaction and energy-deposition simulations based on Monte Carlo codes such as FLUKA [17] and Geant4 [18]. More recently, integrated frameworks such as Xsuite [19] are being developed at CERN to provide a unified and flexible environment for advanced beam dynamics and collimation studies.

### 3.2   Material considerations and mechanical design

To ensure robust operation under the extreme beam conditions of high-power colliders such as the LHC, collimator jaws must withstand high thermal loads, intense radiation fields, and accidental direct beam impacts. Material selection therefore requires a careful balance between high damage resistance, low activation, good thermal conductivity, and sufficient electrical conductivity to limit beam coupling impedance. Carbon-based composites and advanced metal alloys, often complemented by conductive surface coatings, are employed at different collimation stages. From a mechanical perspective, the design must ensure micrometre-level positioning accuracy, long-term dimensional stability, and reliable operation in ultra-high-vacuum and high-radiation environments. Stringent requirements are also imposed on jaw surface flatness and finish to control impedance and preserve beam stability. For reference, details on the LHC collimation system mechanical design can be found in Refs. [20-22].

Figure 5 shows an example of an LHC TCP collimator, consisting of two independently movable jaws that allow precise alignment with respect to the beam. More recent designs, incorporate button beam position monitors (BPMs) at both extremities [23], significantly improving alignment procedures and enabling continuous monitoring of the beam orbit at the collimator location.

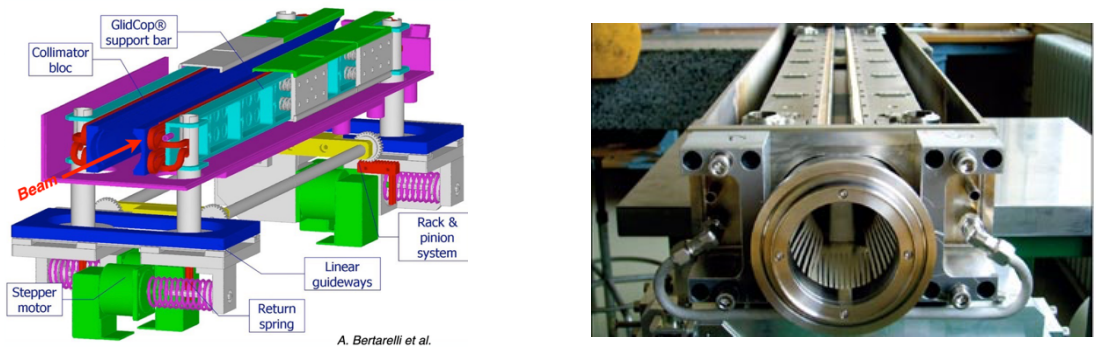

**Fig. 5:** LHC collimator [3]. Left: Schematic view of the LHC primary collimator mechanical assembly. Right: LHC primary collimator illustrating the two independently movable jaws.

### 4   Collimation operational challenges in high-power accelerators

The setup of the collimation systems in high-power accelerators constitutes a critical phase of machine commissioning, as it directly affects machine protection. At the LHC, the beam-based setup procedure begins with aperture measurements aimed at identifying the horizontal and vertical bottleneck for each beam, defining the minimum apertures that must be safeguarded. It then involves precise beam-based alignment of the collimators and the establishment of well-defined operational settings to ensure a robust collimation hierarchy throughout the different stages of the machine cycle. Finally, system performance must be validated through dedicated loss map measurements at low intensity before higher beam intensities can be injected.

## 4.1 Aperture bottleneck measurements

Aperture measurements are aimed at identifying the minimum normalized aperture, commonly referred to as the *bottleneck*, of the machine. While one may rely on the optics model to adjust the collimator settings with predefined safety margins, determining the realistic value of this parameter allows the collimation hierarchy to be tightened in a controlled manner. These measurements have proven particularly valuable for LHC operations, where the typical strategy to meet luminosity targets involved progressively increasing the beam intensity each year. However, in 2017, a further luminosity push required a different strategy due to beam dynamics constraints: specifically, reducing the beam size at the particle physics experiments interaction points. This approach imposed stringent constraints on the minimum aperture that must be protected by the collimation system, since squeezing the beam at the interaction point leads to an increase of the beam size in the final-focus magnets. New squeezed optics configurations were therefore developed to enhance luminosity, and aperture measurements played a key role in validating and pushing forward these new operational modes ensuring safe operation [24].

Several techniques exist to identify the bottleneck, all of which require the use of low-intensity beams, as these measurements deliberately probe the limiting aperture by inducing controlled beam losses. One of the most widely used methods, and still routinely applied in the LHC, is based on inducing a controlled transverse beam blow-up using the transverse damper system until beam losses are observed on a reference collimator with a well-characterized and precisely calibrated normalized aperture. The reference collimator is then progressively opened in small steps until losses are detected at another location in the machine, indicating the presence of a tighter aperture elsewhere. To reliably detect these losses, a comprehensive system of beam loss and radiation monitors distributed around the entire machine is required, with particular emphasis on critical components that, due to the machine optics and component geometry, are potential bottlenecks. These measurements are essential and must therefore be repeated after each technical stop, as the minimum aperture can change due to orbit variations, realignment campaigns, or hardware interventions. A detailed overview of the different techniques developed at CERN can be found in Ref. [14].

## 4.2 Beam-based alignment

Beam-based alignment (BBA) is an essential technique used to align collimators with respect to the beam, ensuring proper setup and optimal performance. This procedure becomes particularly critical in high-power machines with small beam sizes requiring narrow collimator gaps. In the LHC at top energy, beam sizes are on the order of 200 µm, while orbit offsets at the collimators can reach several millimetres and mechanical alignment uncertainties amount to a few hundred micrometres. Under these conditions, the determination of the optimal jaw positions can only be achieved through dedicated beam-based measurements.

The procedure involves shaping the beam halo with a reference collimator set to a known normalized half-gap, then moving other collimator jaws in small steps until symmetric beam losses are observed, providing a direct measurement of the local beam orbit. By iteratively adjusting the reference collimator, the relative beam size and normalized gap settings of each collimator are cross-calibrated, establishing the operational collimation hierarchy based on measured parameters rather than nominal values [25, 26]. At the LHC, due to the procedure's critical importance and complexity, beam-based alignment has evolved from a fully manual procedure to a fully automated one supported by machine learning [27–29]. Given the large number of collimators installed, these advancements have reduced the commissioning time from approximately 20 hours with manual methods to less than one hour with machine learning–assisted methods.

### 4.3 Operational collimator settings generation

In circular high-power colliders such as the LHC, the machine progresses through several operational phases, from injection to top energy and final collision. During this cycle, both beam energy and optics evolve, making the definition of collimator settings challenging. Safe operation requires a fully adjustable, actively controlled collimation system with independently movable jaws that follow continuous, time-dependent functions scaled with energy and optics to preserve the collimation hierarchy. In the LHC, maintaining this hierarchy demands microsecond-level synchronization with the machine timing system to avoid transient bottlenecks or uncontrolled beam losses. This synchronization is one of the most challenging aspects of operation. The detailed definition of the setting functions is beyond the scope of this course (see Ref. [3]). Figure 6 illustrates a typical LHC operational cycle and the corresponding evolution of collimator settings. The top panel shows the currents of selected quadrupoles and dipoles, highlighting the main phases: injection, energy ramp, flat-top, betatron squeeze, and stable beams. The bottom panel displays the half-gaps of selected collimators for beam 1, with TCP, TCSG, and TCT denoting primary, secondary, and tertiary collimators, respectively. At injection, collimators are set to larger gaps to accommodate the large beam size. During the energy ramp, as the beam shrinks, the gaps are progressively reduced following predefined functions that preserve the collimation hierarchy, based on interpolated reference measurements and optics-dependent corrections. The betatron squeeze further narrows the gaps in the experimental regions, and the collimators reach their tightest configuration by the end of the cycle (stable beams).

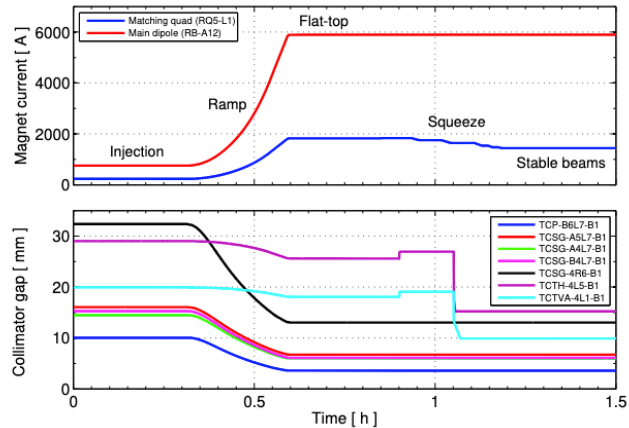

**Fig. 6:** Typical operational cycle of an LHC fill [3]. Top: Evolution of magnet currents. Bottom: Evolution of selected collimator half-gaps during a typical LHC operational cycle, illustrating the synchronized adjustment of collimator settings throughout injection, energy ramp, squeeze, and stable beams.

### 4.4 Loss map validation

Before injecting nominal-intensity beams into a high-power machine, and after the collimation system settings have been implemented, system performance must be validated through dedicated beam-based measurements with low-intensity beams. While internal checks and simulations are important, they cannot fully capture realistic loss conditions and direct measurements are required. This is done by deliberately increasing beam loss rates in a controlled way, for example by exciting the beam with a transverse damper to simulate operational loss scenarios. Loss maps must be measured for each relevant mode to ensure that no unexpected loss peaks occur, that losses in superconducting magnets remain below quench limits, and that the multi-stage collimation hierarchy is preserved. Figure 7 shows a typical LHC loss map: the left plot displays local cleaning inefficiency around the ring, highlighting dominant losses in the collimation insertions, while the right plot zooms into the LHC betatron collimation region (IR7), showing the maintained hierarchy and achieved cleaning efficiency of ~99.993% at the superconducting dipoles.

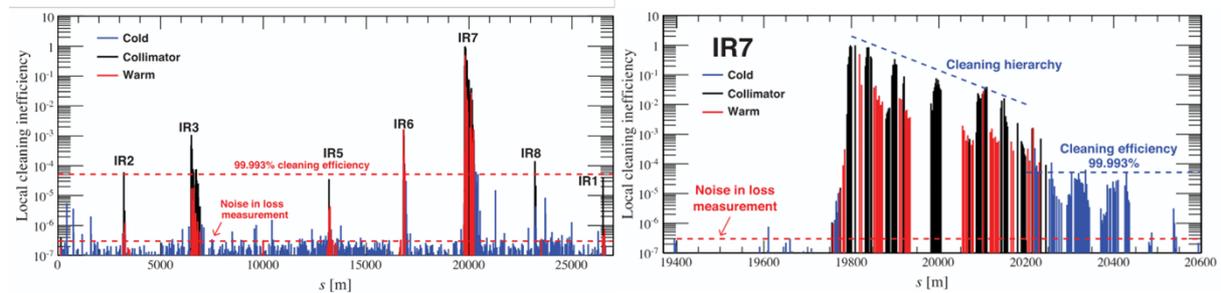

**Fig. 7:** Typical loss map measurement in the LHC [3]. Left: Local cleaning inefficiency around the LHC ring, identifying the dominant loss regions associated with the collimation insertions. Right: Local cleaning inefficiency in the betatron collimation region (IR7).

## 5   R&D on advanced collimation techniques

R&D on advanced collimation techniques continues to explore methods for improved halo control and cleaning efficiency beyond conventional multi-stage systems. One of the most mature approaches is crystal collimation [30], extensively studied since the late 2000s, notably in CERN's UA9 experiment at the SPS, and now operational in the LHC for heavy-ion beams, where bent crystals channel halo particles onto secondary absorbers to reduce losses in sensitive regions. Hollow electron lenses [31] offer another promising concept, using a tuneable annular electron beam to manipulate halo particles while leaving the beam core largely unaffected; developed from Tevatron studies, they have been included in HL-LHC design considerations. Emerging techniques such as tail folding and nonlinear collimation [32] are also under investigation to reshape the halo and mitigate peak losses without perturbing the core. A recent example is the nonlinear collimation system implemented at SuperKEKB in Japan to address beam loss and background challenges, with ongoing studies evaluating its impact on dynamic aperture and beam backgrounds. These innovative approaches, at different stages of maturity, offer potential improvements for future high-intensity and high-power accelerators.

## 6   Concluding remarks

Collimation systems are indispensable in modern high-power accelerators, where they ensure the safe control of both routine and abnormal beam losses while enabling efficient, high-performance operation. This lecture reviewed the fundamental roles of a collimation system, along with the key input parameters that drive its design, including beam size and halo definition, normalized aperture and bottleneck concepts, beam loss mechanisms, and cleaning inefficiency.

The design of a collimation system is governed by several critical factors, encompassing not only performance objectives but also robustness requirements imposed by stored beam energy and beam-dynamics effects such as impedance. These elements must be integrated into a coherent design framework capable of ensuring both reliable machine protection and optimal operational performance. Emphasis was placed on the necessity of a multi-stage collimation layout in high-power machines. Using the LHC as a case study, it was shown that achieving extremely stringent cleaning inefficiencies—at the level of ~0.0001—requires a carefully optimized multi-stage system.

Beyond conceptual and beam dynamics considerations, modern high-power accelerators also impose demanding mechanical constraints. Collimator systems must meet strict requirements for positioning accuracy, manufacturing tolerances, and long-term mechanical stability. Operation is equally sophisticated, involving an extensive program of commissioning measurements to validate settings, hierarchy, and overall system performance. The main procedures implemented at the LHC were briefly described.

The LHC collimation system represents an unprecedented level of complexity in both design and operation. The knowledge and experience gained through its design, commissioning, and operation provide a solid foundation for future high-power facilities. At the same time, ongoing R&D—including crystal collimation, hollow electron lenses, and nonlinear collimation concepts— are being actively pursued to further enhance the performance, robustness, and flexibility of collimation systems for next-generation machines.

**Acknowledgement**

I would like to sincerely thank S. Redaelli for kindly sharing his Collimation CAS 2022 material, which contributed to the preparation of this lecture, as well as the members of the LHC collimation group with whom I have had the opportunity to work and from whom I have learned about collimation systems for high-power hadron colliders.